# Excitons in the wurtzite AlGaN/GaN quantum-well heterostructures

E. P. Pokatilov[a], D. L. Nika[a], V. M. Fomin[b,c]* and J. T. Devreese[b,c]**

[a]*Laboratory of Physics of Multilayer Structures, Department of Theoretical Physics,*

*State University of Moldova, MD-2009 Chişinău, Moldova*

[b]*Theoretische Fysica van de Vaste Stoffen (TFVS), Departement Fysica,*

*Universiteit Antwerpen, B-2020 Antwerpen, Belgium*

[c]*Photonics and Semiconductor Nanostructures (PSN), TU Eindhoven,*

*NL-5600 MB Eindhoven, The Netherlands*

**Abstract**

We have theoretically studied exciton states and photoluminescence spectra of strained wurtzite $Al_xGa_{1-x}N$/GaN quantum-well heterostructures. The electron and hole energy spectra are obtained by numerically solving the Schrödinger equation, both for a single-band Hamiltonian and for a non-symmetrical 6-band Hamiltonian. The deformation potential and spin-orbit interaction are taken into account. For increasing built-in field, generated by the piezoelectric polarization and by the spontaneous polarization, the energy of size quantization rises and the number of size quantized electron and hole levels in a quantum well decreases. The exciton energy spectrum is obtained using electron and hole wave functions and two-dimensional Coulomb wave functions as a basis. We have calculated the exciton oscillator strengths and identified the exciton states active in optical absorption. For different values of the Al content $x$, a quantitative interpretation, in a good agreement with experiment, is provided for (i) the red shift of the zero-phonon photoluminescence peaks for increasing the quantum-well width, (ii) the relative intensities of the zero-phonon and one-phonon photoluminescence peaks, found within the non-adiabatic approach, and (iii) the values of the photoluminescence decay time as a function of the quantum-well width.

---
* Permanent address: Laboratory of Physics of Multilayer Structures, Department of Theoretical Physics, State University of Moldova, MD-2009 Chişinău, Moldova

** E-mail: jozef.devreese@ua.ac.be.

**1. Introduction**

Wurtzite heterostructures with GaN quantum wells (QWs) have a significant potential for electronic and optical applications since GaN possesses a large direct band gap, allows for high temperature stability and high field stability, and is chemically inert. Such heterostructures offer superior characteristics as light emitters in the spectral range from green to ultraviolet [1-3] and as ultrafast optical switches [4, 5]. Experimental and theoretical studies of the nitride heterostructures have attracted much attention [6-14]. Recently, considerable effort has been devoted to investigations of wurtzite $Al_xGa_{1-x}N$/GaN heterostructures with a GaN QW and an $Al_xGa_{1-x}N$ barrier [15-21]. The crystal structure and the electronic properties of the relevant nitride compounds are well established, see e.g. Refs. [22-27].

A large spontaneous electric polarization $P^{SP}$ in the absence of strain and a strong piezoelectric effect characterize wurtzite nitride heterostructures due to the spatial symmetry of their crystal lattice (space group $C_{6v}^4$). Spontaneous $P^{SP}$ and piezoelectric $P^{PZ}$ polarizations generate a built-in electrostatic field, which determines the quantum states of electrons and holes in a GaN QW. This built-in electrostatic field causes peculiarities of the optical absorption- and emission spectra of the wurtzite nitride heterostructures. For instance, with increasing QW width, a red shift of the photoluminescence band (with respect to its position in the bulk crystal) is observed [7, 8, 10, 11, 14, 16, 21]. The influence of the barrier thickness and of the Al concentration in the barrier on the spectral position of the photoluminescence bands was discussed in Refs. [11,14]. Electron intersubband transitions accompanied by absorption and emission of infrared (IR) radiation were detected (Refs. [4,5]). The exciton ground state in a GaN crystal was analyzed, with a degenerate hole band, in the framework of second-order perturbation theory already in Ref. [28]. The influence of the QW width on the exciton ground state energy for a finite-height barrier was investigated using variational methods in Refs. [14, 29]. Experiments on the excitonic absorption and emission in nitride heterostructures were interpreted using variational approaches in Refs. [6, 10, 11, 13, 14, 21]. However, the accuracy of the variational results is restricted.

Another approach consists in the replacement of the multi-band Hamiltonian for holes by a one-band Hamiltonian for light, heavy or spin-orbit split-off holes. Such an approximation is not justified for the calculation of bulk exciton states, because, as demonstrated in Ref. [28], the binding energy of the exciton ground state depends on the effective-mass parameters of the three types of holes. The hole energy spectra, obtained by using a multi-band Hamiltonian for heterostructures [30,31], are not split into the spectra corresponding to the one-band Hamiltonians, except for the particular case when the hole moves only along the $c$-axis and the spin-orbit interaction is neglected.

One-band hole Hamiltonians inevitably lead to fully symmetrical $s$-like ground states for all types of holes (and excitons) confined to spherical quantum dots, whereas holes described by a multi-band Hamiltonian can have a $p$-like ground state [32,33]. The symmetry of the exciton ground state plays a key role for the selection rules and for the interpretation of the absorption and photoluminescence spectra of quantum dots [34]. In a number of experiments (see e. g. Refs. [9,10,12,14]) phonon satellites were detected in photoluminescence spectra of wurtzite nitride heterostructures, providing evidence for the high efficiency of the electron-phonon interaction in GaN QWs.

We have performed a theoretical analysis of the exciton states in wurtzite heterostructures with GaN QWs using a six-band non-symmetrical Hamiltonian for holes and a one-band electron Hamiltonian [30,31]. This approach allows us to describe the exciton energy spectrum and the optical properties of the $Al_xGa_{1-x}N$/GaN wurtzite heterostructures, taking into account the strain due to the mismatch of the crystal lattices of the GaN QW and the $Al_xGa_{1-x}N$ barrier. We then find the energies and the wave functions of the ground state and the excited states of an exciton confined to a QW with the required accuracy, which is controlled by the size of the selected wave-function basis. Treating the problem beyond the framework of the one-band approximation, we take into account the mixing of light, heavy and spin-orbit split-off holes.

This article is organized as follows. In Sec. II we describe the built-in electrostatic field in the $Al_xGa_{1-x}N/GaN$ heterostructures. The electron and hole energy spectra and the wave functions are obtained in Sec. III. The exciton problem is solved in Sec. IV. In Sec. V the oscillator strength and the exciton photoluminescence decay time in the $Al_xGa_{1-x}N/GaN$ heterostructures are described. Results of the calculations are compared with the experimental data in Sec. VI. Section VII contains the conclusions.

**II. Built-in electrostatic field in** $Al_xGa_{1-x}N/GaN$ **QW heterostructures**

Wurtzite heterostructures are usually grown on a GaN substrate by molecular beam epitaxy. The hexagonal reference $c$-axis is oriented along the direction of the crystal growth, perpendicular to the heterostructure interfaces. A multi-quantum well (MQW) heterostructure contains GaN QWs (of width $d_1$) and $Al_xGa_{1-x}N$ barriers (of thickness $d_2$), with $d_2 > d_1$. A Cartesian coordinate system $(X,Y,Z)$ is used with origin in the middle of the GaN QW layer. The $Z$-axis is parallel to the hexagonal reference $c$-axis. The axes $X,Y$ are arbitrarily oriented in the middle plane of the GaN layer.

Since the wurtzite crystal lattices of GaN and AlGaN lack inversion symmetry, (i) the heterostructure layers are spontaneously polarized with the spontaneous polarization $P^{SP}$ oriented along the $c$-axis, and (ii) the strain due to the lattice mismatch between GaN and AlGaN generates the piezoelectric polarization $P^{PZ}$ [35]. Wurtzite crystals have three different piezoelectric coefficients are $e_{15}, e_{31}$ and $e_{33}$. The piezoelectric polarization along the $c$-axis is:

$$P^{PZ} = e_{31}(\varepsilon_{xx} + \varepsilon_{yy}) + e_{33}\varepsilon_{zz}, \tag{1}$$

where $\varepsilon_{lm} = \frac{1}{2}(\frac{\partial U_l}{\partial x_m} + \frac{\partial U_m}{\partial x_l})$ are the components of the strain tensor, $U_l$ is the component of the displacement vector, and the indices run over the spatial coordinates $X$, $Y$ and $Z$.

X-ray diffraction mapping has shown [10,12] that the samples are pseudomorphically strained on the GaN substrate. In other words, the barrier layers of Al$_x$Ga$_{1-x}$N in the heterostructure are deformed, so that the lattice constant is adjusted to the GaN substrate and to the QWs. The components of the strain tensor in the barrier layers are

$$\varepsilon_{xx}^b = \varepsilon_{yy}^b = \frac{a(\text{GaN}) - a(\text{Al}_x\text{Ga}_{1-x}\text{N})}{a(\text{GaN})} \qquad (2)$$

and

$$\varepsilon_{zz}^b = -2\frac{c_{13}^b}{c_{33}^b}\varepsilon_{xx}^b. \qquad (3)$$

As follows from (2) and (3), the piezoelectric polarization in the barrier layers is

$$P_b^{PZ} = 2\varepsilon_{xx}^b(e_{31}^b - e_{33}^b \frac{c_{13}^b}{c_{33}^b}). \qquad (4)$$

The polarity of the spontaneous polarization $P^{SP}$ is specified by the terminating anion or cation at the surface [36,37]. The total polarization

$$\vec{P} = \vec{P}^{PZ} + \vec{P}^{SP} \qquad (5)$$

entails an electrostatic potential in the heterostructure.

The electric displacement vector in each layer is $\vec{D} = \varepsilon_0 \varepsilon \vec{F} + \vec{P}$, where $\vec{F}$ is the electric field, $\varepsilon$ is the dielectric constant and $\varepsilon_0$ is the permittivity of the vacuum. In the absence of the bulk electric charges, it follows from the Maxwell equation, $\text{div}\vec{D} = 0$, that a uniform electrostatic field $F$ exists in the QWs and in the barrier layers. This field is determined by both constituents of the polarization in (5) and it is directed along the hexagonal reference $c$-axis. From the conditions of (i) continuity of the normal component $D_z$ at the interfaces [14] and

(ii) vanishing potential at the external surfaces of the MQW heterostructure, the expression for the electrostatic field in a QW takes the form:

$$F = \frac{L_b F_0}{L_b + \varepsilon_b^s / \varepsilon_w^s L_w} \tag{6}$$

with $L_{w(b)} = N_{w(b)} l_{w(b)}$. $N_{w(b)}$ is the number of QWs (barrier layers) in the MQW heterostructure and $l_{w(b)}$ is the QW width (barrier thickness), $\varepsilon_{w(b)}^s$ is the static dielectric constant of the QW (barrier layer). For unstrained GaN layers $F_0 = \left\| |P_w^{SP}| - |P_b^{SP}| - |P_b^{PZ}| \right\| / \varepsilon_0 \varepsilon_w^s$. The electrostatic field in an $Al_xGa_{1-x}N$ barrier layer is

$$F_b = -\frac{L_w F}{L_b}. \tag{7}$$

For $N_w = N_d \Box 1$, Eqs. (6) and (7) are the results known for a superlattice [14,36]. The spontaneous polarization vectors in both layers were assumed as mutually parallel [37], independently of whether the boundary surfaces are Ga-faced or N-faced. The potential energy of the electron in the GaN QW in the region $-d_1/2 < z < d_1/2$, with the built-in electrostatic field $F$, is

$$V_P(z) = -eFz, \quad -d_1/2 < z < d_1/2, \tag{8}$$

where $e$ is the electron charge.

The built-in field $F$, calculated using the theoretical values of the piezo-moduli and of the spontaneous polarization from Ref. [36], is larger than that derived from the experimental photoluminescence peaks of the wurtzite heterostructures [12]. In Ref. [12] it was shown that the built-in field $F$ depends on the temperature conditions of the sample growth. In samples grown at 650°C [12] a strong field $F_{exp}$ =1300 kV/cm was observed, whereas in samples fabricated at 850°C, $F_{exp}$ is considerably smaller, between 530 kV/cm and 760 kV/cm (depending on the thickness of the $Al_xGa_{1-x}N$ barrier layers) [14]. These fitting values of the built-in field agree within 15-19% with the theoretical values $F_{theor}^{PZ}$ calculated using the piezoelectric polarization only. Consequently, the contribution of the spontaneous polarization to the observed built-in field is considerably smaller than the theoretical value $F_{theor}^{SP}$. A possible reason for the smallness of the spontaneous polarization at higher growth temperatures is the thermo-diffusion of Al [12]. The thermo-diffusion "smoothes out" the step-like distribution of Al at interfaces, and hence, leads to a decrease of the value of the spontaneous polarization in comparison with its theoretical value calculated for an ideal position of the cationic and anionic lattice sites near the interface. Because the spontaneous polarization in the experiment cannot be controlled, the following combination of the polarizations, $P_0 = \left\| \left| P_w^{SP} \right| - \left| P_b^{SP} \right| - \left| P_b^{PZ} \right| \right\|$, is considered as a fitting parameter.

### III. Electron and hole states in wurtzite Al$_x$Ga$_{1-x}$N/GaN heterostructures

The electron wave function is $\phi_e | S >$ with $| S >$, the Bloch wave function corresponding to the conduction-band bottom. Electron states are eigenstates of the Schrödinger equation:

$$\hat{H}_e \phi_e = E^e \phi_e, \tag{9}$$

where $\phi_e$ is the electron envelope wave function. The electron Hamiltonian $\hat{H}_e$ is

$$\hat{H}_e = \hat{H}_s(\vec{r}_e) + V_P(z_e) + H^{(\varepsilon)}(\vec{r}_e) + \Delta E_c(z_e) + V_{SA}(z_e), \tag{10}$$

where $\hat{H}_s(\vec{r}_e)$ is the kinetic part of the Hamiltonian for a unit cell averaged with the Bloch wave function $| S >$ and $V_{SA}(z_e)$ is the electron self-interaction energy, which is known analytically for a three-layer heterostructure (see Ref. [38]). The energy $E_c$(GaN) of the bottom of the conduction band in unstrained GaN is chosen as the reference energy level: $\Delta E_c = 0, | z_e | \leq d_1/2$. The height of the potential barrier is $\Delta E_c = \delta E_c = E_c(\text{Al}_x\text{Ga}_{1-x}\text{N}) - E_c(\text{GaN})$, $| z_e | > d_1/2$, where $\delta E_c$ is the conduction-band offset and $E_c$(Al$_x$Ga$_{1-x}$N) is the energy of the bottom of the conduction band in unstrained Al$_x$Ga$_{1-x}$N.

The strain-dependent part of the electron Hamiltonian (10) is

$$H_e^{(\varepsilon)}(\vec{r}_e) = a_c^{\Box}(\vec{r}_e)\varepsilon_{zz}(\vec{r}_e) + a_c^{\perp}(\vec{r}_e)[\varepsilon_{xx}(\vec{r}_e) + \varepsilon_{yy}(\vec{r}_e)], \tag{11}$$

where $a_c^{\parallel}$ and $a_c^{\perp}$ are the conduction-band deformation potentials [35]. The zero-strain conduction-band and the valence-band offsets are taken from Ref. [39] with the bowing parameter $b=1$ eV. The bowing parameter takes into account a non-linear dependence of the band offsets in $Al_xGa_{1-x}N$ on the aluminum composition. Due to the strong electron confinement, we can consider the electron motion along the $Z$- axis (the size-quantized motion) as "fast" and the motion in the $(X, Y)$-plane as "slow". The Schrödinger equation for the "fast" motion takes the form:

$$\left\{-\frac{\hbar^2}{2}\frac{\partial}{\partial z_e}\left(\frac{1}{m_e^{\parallel}}\right)\frac{\partial}{\partial z_e}+U(z_e)\right\}\phi_l(z_e)=E_l^{e,\parallel}\phi_l(z_e), \quad (12)$$

where $U(z_e)=-eFz_e+H_e^{(\varepsilon)}(z_e)+\Delta E_c(z_e)+V_{SA}(z_e)$. We solve this equation, using a finite-difference method (the numerical error for the obtained energies does not exceed 0.5%). The ground-state electron wave functions are schematically shown in Fig. 1 for a rectangular ($F=0$) and a triangular ($F\neq 0$) potential profiles.

The hole wave function is $\psi_h=(\vec{u},\vec{\Psi}_h)$, where $\vec{u}=(|X\rangle|\uparrow\rangle, |Y\rangle|\uparrow\rangle, |Z\rangle|\uparrow\rangle, |X\rangle|\downarrow\rangle, |Y\rangle|\downarrow\rangle, |Z\rangle|\downarrow\rangle)$, where $|X\rangle, |Y\rangle$ and $|Z\rangle$ are the Bloch wave functions corresponding to the top of the valence band, $|\uparrow\rangle, |\downarrow\rangle$ are the spin functions of the missing electron and $\vec{\Psi}_h$ is a six-component column vector representing the hole envelope wave function. Hole states are eigenstates of the six-band Schrödinger equation:

$$\hat{H}_h\vec{\Psi}_h=E^h\vec{\Psi}_h, \quad (13)$$

where $\hat{H}_h$ is the hole Hamiltonian and $E^h$ is the hole eigenenergy. The hole Hamiltonian includes the spin-orbit interaction, the interaction with the lattice deformation and the interaction with the electrostatic built-in field due to the piezoelectric and the spontaneous polarizations in the heterostructure. The six-band Hamiltonian $\hat{H}_h$ is

$$\hat{H}_h=\begin{vmatrix} \hat{H}_{XYZ}(\vec{r}_h)+\hat{H}_h^{\varepsilon} & 0 \\ 0 & \hat{H}_{XYZ}(\vec{r}_h)+\hat{H}_h^{\varepsilon} \end{vmatrix}+\hat{H}_{S-O}+eFz_h\hat{1}+\Delta E_h(z_h)\hat{1}+V_{SA}(z_h)\hat{1}. \quad (14)$$

$\Delta E_h(z_h)$ is the hole barrier height: $\Delta E_h = -E_g(\text{GaN})$, $|z_h| \leq d_1/2$; $\Delta E_h = -E_g(\text{GaN}) - \delta E_h$, $z_h > |d_1/2|$, $\delta E_h$ is the valence-band offset, $V_{SA}(z_h)$ is the hole self-interaction energy (see Ref. [38] for an analytical representation in the case of a three-layer heterostructure), $\hat{1}$ is the $6 \times 6$ unit matrix, $H_{XYZ}$ is the $3 \times 3$ matrix representing the kinetic energy in the Hamiltonian for a unit cell in the basis $|X>$, $|Y>$, $|Z>$:

$$\hat{H}_{XYZ} = \frac{\hbar^2}{2m_0} \| h_{ik} \|, \qquad (15)$$

where

$$\hat{h}_{11} = \frac{\hbar^2}{2m_0}(\hat{k}_x L_1 \hat{k}_x + \hat{k}_y M_1 \hat{k}_y + \hat{k}_z M_2 \hat{k}_z), \; \hat{h}_{12} = \frac{\hbar^2}{2m_0}(\hat{k}_x N_1 \hat{k}_y + \hat{k}_y N_1' \hat{k}_x),$$

$$\hat{h}_{13} = \frac{\hbar^2}{2m_0}(\hat{k}_x N_2 \hat{k}_z + \hat{k}_z N_2' \hat{k}_x), \; \hat{h}_{21} = \frac{\hbar^2}{2m_0}(\hat{k}_x N_2 \hat{k}_z + \hat{k}_z N_2' \hat{k}_x),$$

$$\hat{h}_{22} = \frac{\hbar^2}{2m_0}(\hat{k}_x M_1 \hat{k}_x + \hat{k}_y L_1 \hat{k}_y + \hat{k}_z M_2 \hat{k}_z), \; \hat{h}_{23} = \frac{\hbar^2}{2m_0}(\hat{k}_y N_2 \hat{k}_z + \hat{k}_z N_2' \hat{k}_y), \qquad (16)$$

$$\hat{h}_{31} = \frac{\hbar^2}{2m_0}(\hat{k}_z N_2 \hat{k}_x + \hat{k}_x N_2' \hat{k}_z), \; \hat{h}_{32} = \frac{\hbar^2}{2m_0}(\hat{k}_z N_2 \hat{k}_y + \hat{k}_y N_2' \hat{k}_z),$$

$$\hat{h}_{33} = \frac{\hbar^2}{2m_0}(\hat{k}_x M_3 \hat{k}_x + \hat{k}_y M_3 \hat{k}_y + \hat{k}_z L_2 \hat{k}_z - \delta_{cr})$$

with $L_1 = A_2 + A_4 + A_5$, $L_2 = A_1$, $M_1 = A_2 + A_4 - A_5$, $M_2 = A_1 + A_3$, $M_3 = A_2$, $N_1 = 3A_5 - (A_2 + A_4) + 1$, $N_1' = -A_5 + A_2 + A_4 - 1$, $N_2 = 1 - (A_1 + A_3) + \sqrt{2}A_6$, $N_2' = A_1 + A_3 - 1$ and the parameter $\delta_{cr}$ determines the crystal-field splitting energy $-\hbar^2 \delta_{cr}/(2m_0)$. The coefficients $A_i$ ($i=1,2,3$) are the Rashba-Sheka-Pikus parameters of the valence band [35]

$$\begin{aligned} H_h^\varepsilon &= |d_{lm}|, \\ d_{11} &= l_1 \varepsilon_{xx} + m_1 \varepsilon_{yy} + m_2 \varepsilon_{zz}, \; d_{12} = n_1 \varepsilon_{xy}, \; d_{13} = n_2 \varepsilon_{xz}, \\ d_{22} &= m_1 \varepsilon_{xx} + l_1 \varepsilon_{yy} + m_2 \varepsilon_{zz}, \; d_{23} = n_2 \varepsilon_{yz}, \\ d_{33} &= m_3 (\varepsilon_{xx} + \varepsilon_{yy}) + l_2 \varepsilon_{zz}, \end{aligned} \qquad (17)$$

where

$$l_1 = D_2 + D_4 + D_5, \ l_2 = D_1,$$
$$m_1 = D_2 + D_4 - D_5, \ m_2 = D_2 + D_3, \ m_3 = D_2, \quad (18)$$
$$n_1 = 2D_5, \ n_2 = \sqrt{2}D_6.$$

The coefficients $D_k$ ($k=1,…,6$) are the valence-band deformation potentials [35]. The nonsymmetrical three-band Hamiltonian $\hat{H}_{XYZ}$ (15) and the deformation interaction Hamiltonian $H_h^\varepsilon$ (17) were derived in Ref. [30]. The spin–orbit Hamiltonian [30,31] is

$$\hat{H}_{S-O}(\vec{r}_h) = A_{SO}(\vec{r}_h)\begin{pmatrix} -1 & -i & 0 & 0 & 0 & 1 \\ i & -1 & 0 & 0 & 0 & -i \\ 0 & 0 & -1 & -1 & i & 0 \\ 0 & 0 & -1 & -1 & i & 0 \\ 0 & 0 & -i & -i & -1 & 0 \\ 1 & i & 0 & 0 & 0 & -1 \end{pmatrix}. \quad (19)$$

In Eq. (19), $A_{SO}(\vec{r}_h) = \Delta_{S-O}(\vec{r}_h)/3$, where $\Delta_{SO}(\vec{r}_h)$ is the spin-orbit splitting energy. Finally, the hole Hamiltonian (14) can be represented as the $6\times 6$ matrix:

$$\hat{H}_h = \hat{H}_{h1} + \hat{H}_{h2}, \quad (20)$$

$$\hat{H}_{h1} = \begin{pmatrix} \hat{h}_{11} - A_{SO} + d_{11} & \hat{h}_{12} - iA_{SO} + d_{12} & \hat{h}_{13} + d_{13} & 0 & 0 & A_{SO} \\ \hat{h}_{21} + iA_{SO} + d_{12} & \hat{h}_{22} - A_{SO} + d_{22} & \hat{h}_{23} + d_{23} & 0 & 0 & -iA_{SO} \\ \hat{h}_{31} + d_{13} & \hat{h}_{32} + d_{23} & \hat{h}_{33} - A_{SO} + d_{33} & -A_{SO} & iA_{SO} & 0 \\ 0 & 0 & -A_{SO} & \hat{h}_{11} - A_{SO} + d_{11} & \hat{h}_{21} + iA_{SO} + d_{12} & \hat{h}_{13} + d_{13} \\ 0 & 0 & -iA_{SO} & \hat{h}_{21} - iA_{SO} + d_{12} & \hat{h}_{22} - A_{SO} + d_{22} & \hat{h}_{23} + d_{23} \\ A_{SO} & iA_{SO} & 0 & \hat{h}_{31} + d_{13} & \hat{h}_{32} + d_{23} & \hat{h}_{33} - A_{SO} + d_{33} \end{pmatrix},$$

$$\hat{H}_{h2} = \left[eFz_h + \Delta E_h(z_h) + V_{SA}(z_h)\right]\hat{1}.$$

The values of the material parameters, required for numerical calculations, are taken from Ref. [35].

The hole size-quantized wave function $\vec{\Upsilon}(z_h)$ with six components $\Upsilon_i(z_h)$, $i=1$ to 6, are found from Eq. (13) with the Hamiltonian (20) at $k_x = k_y = 0$. The six-band Schrödinger equation (13) is a set of 6 algebraic equations, which splits into two independent sets of 3 equations: for the components ($\Upsilon_1, \Upsilon_2, \Upsilon_4$) and ($\Upsilon_3, \Upsilon_5, \Upsilon_6$). These sets of equations are solved numerically, using a finite-difference method. A scheme of the heterostructure and of the electron and hole energy levels is presented in Fig. 2. The number of size-quantized electron and hole energy levels in the QW depends on the electric field. The obtained electron and hole states are used in the next section to solve the exciton problem.

**IV. Exciton states in wurtzite $Al_xGa_{1-x}N/GaN$ heterostructures**

The exciton Hamiltonian for the heterostructure under consideration

$$\hat{H}_{exc} = \hat{H}_e(\vec{r}_e) + \hat{H}_h(\vec{r}_h) + V_C(\vec{r}_e - \vec{r}_h, z_e, z_h)\hat{1}, \quad \vec{r} = (x, y) \quad (21)$$

includes the electron one-band Hamiltonian (10), the hole six-band Hamiltonian (20), and the potential energy of the electron-hole Coulomb interaction $V_C(\vec{r}_e - \vec{r}_h, z_e, z_h)$.

The potential energy of the electron-hole Coulomb interaction $V_C(\vec{r}_e - \vec{r}_h, z_e, z_h)$ for a three-layer heterostructure has been derived analytically in Ref. [38]. When both charge carriers are in the QW, it takes the form:

$$\begin{aligned}
V_C(\vec{r}_e - \vec{r}_h, |z_e - z_h|) &= -\frac{e^2}{4\pi\varepsilon_0\varepsilon_w}[\frac{1}{\sqrt{(z_e - z_h)^2 + \rho^2}} + 2\delta_\varepsilon B_1 + 2\delta_\varepsilon^2 B_2], \\
B_1 &= \int_0^\infty \frac{\exp[-\eta d_1]\cosh[\eta(z_e + z_h)]}{1 - \delta_\varepsilon^2 \exp[-2\eta d_1]} J_0(\eta\rho)d\eta, \\
B_2 &= \int_0^\infty \frac{\exp[-2\eta d_1]\cosh[\eta(z_e - z_h)]}{1 - \delta_\varepsilon^2 \exp[-2\eta d_1]} J_0(\eta\rho)d\eta, \\
\delta_\varepsilon &= \frac{\varepsilon_w - \varepsilon_b}{\varepsilon_w + \varepsilon_b}, \rho^2 = (x_e - x_h)^2 + (y_e - y_h)^2,
\end{aligned} \quad (22)$$

where $\varepsilon_{w(b)}$ is the optical dielectric constant of the QW (barrier layer), $J_0$ is the Bessel function of the first kind of index zero [40].

To study localized exciton states, we use the coordinates

$$x = x_e - x_h, \quad y = y_e - y_h, \quad (23)$$

and the momenta

$$k_{ex} = k_x; \quad k_{ey} = k_y \text{ and } k_{hx} = -k_x; \quad k_{hy} = -k_y. \quad (24)$$

These relations between the electron and hole momenta follow from the condition $\vec{K} = \vec{k}_e + \vec{k}_h = 0$, where $\vec{K}$ is the exciton momentum. The variables given by (23), (24) are convenient for the calculation of the internal states of the exciton. The exciton energies and wave functions are eigenenergies and eigenfunctions of the six-band envelope-function Schrödinger equation:

$$\hat{H}_{exc}(x, y, z_e, z_h)\Psi_{exc}(x, y, z_e, z_h) = E_{exc}\Psi_{exc}(x, y, z_e, z_h). \qquad (25)$$

To take into account the size-quantized motion of both charge carriers, the Hamiltonian $\hat{H}_{exc}$ is averaged using the product of the electron and hole wave functions $\phi_e^i(z_e)\vec{\Upsilon}^j(z_h)$:

$$H_{exc}^{ij}(x,y) = \int \phi_e^{i*}(z_e)(\vec{\Upsilon}^j)^+(z_h)\{\hat{H}_e(\vec{r}_e) + \hat{H}_h(\vec{r}_h) + V_C(\vec{r}_e - \vec{r}_h, |z_e - z_h|)\hat{1}\}\phi_e^i(z_e)\vec{\Upsilon}^j(z_h)dz_e dz_h =$$

$$E_e^i + E_h^j - \frac{\hbar^2}{2m_{xx}^{ij}}\frac{\partial^2}{\partial x^2} - \frac{\hbar^2}{2m_{yy}^{ij}}\frac{\partial^2}{\partial y^2} - \frac{\hbar^2}{2m_{xy}^{ij}}\frac{\partial}{\partial x}\frac{\partial}{\partial y} + \overline{V}_{Coulomb}(x,y), \qquad (26)$$

where $E_e^i$ is the energy of the size-quantized electron state with quantum number $i$ ($i = 1,...,I$), $E_j^h$ is the energy of the size-quantized hole state with quantum number $j$ ($j = 1,...,J$), and the following notations are used:

$$\frac{1}{m_{xx}^{ij}} = \int_{-\frac{L}{2}}^{\frac{L}{2}} |\phi_e^i(z_e)|^2 \frac{1}{m_\perp(z_e)}dz_e -$$

$$\int_{-\frac{L}{2}}^{\frac{L}{2}} \{L_1((\Upsilon_1^j)^2 + (\Upsilon_2^j)^2) + M_1((\Upsilon_3^j)^2 + (\Upsilon_4^j)^2) + M_3((\Upsilon_5^j)^2 + (\Upsilon_6^j)^2)\}dz_h, \qquad (27)$$

$$\frac{1}{m_{yy}^{ij}} = \int_{-\frac{L}{2}}^{\frac{L}{2}} |\phi_e^i(z_e)|^2 \frac{1}{m_\perp(z_e)}dz_e -$$

$$\int_{-\frac{L}{2}}^{\frac{L}{2}} \{M_1((\Upsilon_1^j)^2 + (\Upsilon_2^j)^2) + L_1((\Upsilon_3^j)^2 + (\Upsilon_4^j)^2) + M_3((\Upsilon_5^j)^2 + (\Upsilon_6^j)^2)\}dz_h, \qquad (28)$$

$$\frac{1}{m_{xy}^{ij}} = -2 \int_{-\frac{L}{2}}^{\frac{L}{2}} \{ \Upsilon_1^j (N_1 + N_1') \Upsilon_3^j + \Upsilon_2^j (N_1 + N_1') \Upsilon_4^j \} dz_h, \tag{29}$$

$$\overline{V}_{Coulomb}^{ij}(x,y) = \int_{-\frac{L}{2}}^{\frac{L}{2}} \int_{-\frac{L}{2}}^{\frac{L}{2}} (|\phi_e^i(z_e)|^2 |\Upsilon^j(z_h)|^2 V_C(x,y,|z_e - z_h|)) dz_e dz_h. \tag{30}$$

Next, for every pair of the indices $i,j$, we numerically solve the Schrödinger equation with the Hamiltonian (26):

$$H_{exc}^{ij} \Phi_k^{ij}(x,y) = E_k^{ij} \Phi_k^{ij}(x,y) \tag{31}$$

to find the exciton Coulomb functions $\Phi_k^{ij}(x,y)$ ($k = 1,...,K$).

The set of product functions $\phi_e^i(z_e) \vec{\Upsilon}^j(z_h) \Phi_k^{ij}(x,y)$ forms an orthonormalized basis:

$$\int (\phi_e^i(z_e) \vec{\Upsilon}^j(z_h) \Phi_k^{ij}(x,y))^+ (\phi_e^{i'} \vec{\Upsilon}^{j'}(z_h) \Phi_{k'}^{i'j'}(x,y)) dz_e dz_h dx dy = \delta_{kk'} \delta_{jj'} \delta_{ii'}. \tag{32}$$

The exciton wave functions are then given as the following expansion:

$$\Psi_{exc}^{n,\alpha}(x, y, z_e, z_h) = \sum_{i,j,k} C_{i,j,k}^{n,\alpha} \phi_e^i(z_e) \vec{\Upsilon}^j(z_h) \Phi_k^{ij}(x,y). \tag{33}$$

Here, $n$ is a quantum number of an exciton state, $\alpha$ is the degree of degeneracy of the exciton state with quantum number $n$ and $C_{i,j,k}^{n,\alpha}$ are expansion coefficients. In the rhs of Eq. (33), *IJK* basis functions are used: *I* size-quantized electronic functions ($i = 1,...,I$), *J* size-quantized hole functions ($j = 1,...,J$) and – for every pair of indices $i,j$ – *K* exciton Coulomb functions ($k = 1,...,K$).

Next, Eq. (25) is projected onto the selected basis:

$$\int (\varphi_e^i \vec{\Upsilon}^j \Phi_k^{ij})^+ \hat{H}_{exc} \Psi_{exc}^{n,\alpha} dz_e dz_h dx dy = E_{exc}^{n,\alpha} \int (\varphi_e^i \vec{\Upsilon}^j \Phi_k^{ij})^+ \Psi_{exc}^{n,\alpha} dz_e dz_h dx dy, \tag{34}$$

resulting in the the set of *IJK* equations:

$$\sum_{i',j',k'} \left\{ (E_e^i + E_h^j + E_g)\delta_{i,i'}\delta_{j,j'}\delta_{k,k'} + \frac{\hbar^2}{2\overline{m}_\perp^{i,i'}}(F_{xx}^{i,i',j,j',k,k'} + F_{yy}^{i,i',j,j',k,k'})\delta_{i,i',j,j'} \right.$$
$$\left. -\left(\frac{\hbar^2}{2\overline{m}_{xx}^{j,j'}}F_{xx}^{i,i',j,j',k,k'} + \frac{\hbar^2}{2\overline{m}_{yy}^{j,j'}}F_{yy}^{i,i',j,j',k,k'} + \frac{\hbar^2}{2\overline{m}_{xy}^{j,j'}}F_{xy}^{i,i',j,j',k,k'}\right)\delta_{i,i'} + V_{Coulomb}^{j,j',k,k'} \right\} C_{i',j',k'}^{n,\alpha} = E_{exc}^{n,\alpha} C_{i,j,k}^{n,\alpha}. \quad (35)$$

Here

$$\frac{1}{\overline{m}_\perp^{i,i'}} = \int_{-\infty}^{\infty} \phi_e^{i'*}(z_e) \frac{1}{m_\perp(z_e)} \phi_e^i(z_e) dz_e, \quad (36)$$

$$\frac{1}{\overline{m}_{xx}^{j,j'}} = \int_{-\infty}^{\infty} (L_1(z_h)\{\Upsilon_1^{j'}\Upsilon_1^j + \Upsilon_2^{j'}\Upsilon_2^j\} + M_1(z_h)\{\Upsilon_3^{j'}\Upsilon_3^j + \Upsilon_4^{j'}\Upsilon_4^j\} + M_3(z_h)\{\Upsilon_5^{j'}\Upsilon_5^j + \Upsilon_6^{j'}\Upsilon_6^j\}) dz_h, \quad (37)$$

$$\frac{1}{\overline{m}_{yy}^{j,j'}} = \int_{-\infty}^{\infty} (M_1(z_h)\{\Upsilon_1^{j'}\Upsilon_1^j + \Upsilon_2^{j'}\Upsilon_2^j\} + L_1(z_h)\{\Upsilon_3^{j'}\Upsilon_3^j + \Upsilon_4^{j'}\Upsilon_4^j\} + M_3(z_h)\{\Upsilon_5^{j'}\Upsilon_5^j + \Upsilon_6^{j'}\Upsilon_6^j\}) dz_h, \quad (38)$$

$$\frac{1}{\overline{m}_{xy}^{j,j'}} = \int_{-\infty}^{\infty} (N_1(z_h) + N_1^{'}(z_h)\{\Upsilon_1^{j'}\Upsilon_3^j + \Upsilon_2^{j'}\Upsilon_4^j + \Upsilon_3^{j'}\Upsilon_1^j + \Upsilon_4^{j'}\Upsilon_2^j\}) dz_h, \quad (39)$$

$$F_{xx}^{j,j',k,k'} = \int_{-\infty}^{\infty}\int_{-\infty}^{\infty} \frac{\partial \Phi_{k'}^{j'}}{\partial x}\frac{\partial \Phi_k^j}{\partial x} dxdy, \quad F_{yy}^{j,j',k,k'} = \int_{-\infty}^{\infty}\int_{-\infty}^{\infty} \frac{\partial \Phi_{k'}^{j'}}{\partial y}\frac{\partial \Phi_k^j}{\partial y} dxdy,$$
$$F_{xy}^{i,i',j,j',k,k'} = \int_{-\infty}^{\infty}\int_{-\infty}^{\infty} \frac{\partial \Phi_{k'}^{j'}}{\partial x}\frac{\partial \Phi_k^j}{\partial y} dxdy. \quad (40)$$

Finally, we obtain the exciton energy spectrum $E_{exc}^{n,\alpha}$ and the coefficients $C_{i,j,k}^{n,\alpha}$ numerically from the set (35).

### V. Oscillator strengths. Photoluminescence and radiation decay time

The oscillator strength of an exciton state with quantum number $n$ within the effective mass approach is

$$f_n = \frac{2\hbar^2}{m_0 E_{exc}^n} \sum_\alpha |\int dz_e dz_h dxdy \delta(z_e - z_h)\delta(x)\delta(y)(\vec{e},\vec{k}_h)\Psi_{exc}^{n,\alpha}|^2 \quad (41)$$

with $\vec{e}$ the polarization vector of the incident light. When the incident light is polarized in the plane perpendicular to the $c$-axis, $(\vec{e},\vec{k}_h) = e_x k_{h,x} + e_y k_{h,y}$, and the oscillator strength (41) is

$$f_n^{\perp} = \frac{E_P}{E_{exc}^n} \sum_{\alpha} \left\{ (\sum_{i,j,k} C_{i,j,k}^{n,\alpha} \Phi_k^{ij}(0,0) \int_{-\infty}^{\infty} \phi_e^i(z) \Upsilon_1^j(z)dz)^2 + (\sum_{j,k} C_{i,j,k}^{n,\alpha} \Phi_k^{ij}(0,0) \int_{-\infty}^{\infty} \phi_e^i(z) \Upsilon_2^j(z)dz)^2 \right\}. \quad (42)$$

For light polarized along the $c$-axis, $(\vec{e},\vec{k}_h) = e_z k_{h,z}$, and the oscillator strength (41) has the form

$$f_n^{\square} = \frac{E_P}{E_{exc}^n} \sum_{\alpha} \left\{ (\sum_{i,j,k} C_{i,j,k}^{n,\alpha} \Phi_k^{ij}(0,0) \int_{-\infty}^{\infty} \phi_e^i(z) \Upsilon_5^j(z)dz)^2 + (\sum_{i,j,k} C_{i,j,k}^{n,\alpha} \Phi_k^{ij}(0,0) \int_{-\infty}^{\infty} \phi_e^i(z) \Upsilon_6^j(z)dz)^2 \right\}. \quad (43)$$

In (42) and (43), $E_P$ is the Kane energy, determined from the definition:

$$<S|k_i|I> = \delta_{i,I}\sqrt{\frac{m_0 E_P}{2\hbar^2}}, \quad (44)$$

where $|I> = |X>, |Y>, |Z>$.

In the low-temperature limit and in the dipole approximation, the intensities of the $N$-phonon lines in the photoluminescence spectrum using the non-adiabatic approach have been obtained in Ref. [41].

$$I(\Omega) \sim \sum_{\lambda,N} F_{n_0,N}^{(\lambda)} \delta(\Omega_{n_0} - N\omega_{\lambda} - \Omega), \quad (45)$$

where $I(\Omega)$ is the luminescence intensity, $\Omega$ is the frequency of the emitted light, $\Omega_{n_0} = E_{n_0}/\hbar$, $E_{n_0}$ is the exciton ground-state energy, $\omega_{\lambda}$ is the frequency of the $\lambda$ th phonon mode. The amplitude $F_{n_0,N}^{(\lambda)}$ in (45) depends on the number $N$ of phonons ($N=0, 1,\ldots$) participating in the photoluminescence process: for the zero-phonon line

$$F_{n_0,0} = |f_{n_0}|^2 \quad (46)$$

and for the one-phonon line

$$F_{n_0,1}^{(\lambda)} = \sum_{n_1,n_2} \frac{f_{n_1}^* f_{n_2} <n_1|\gamma_{\lambda}|n_0><n_0|\gamma_{\lambda}^*|n_2>}{\hbar^2(\Omega_{n_0} - \Omega_{n_1} - \omega_{\lambda})(\Omega_{n_0} - \Omega_{n_2} - \omega_{\lambda})}, \quad (47)$$

where $\gamma_\lambda$ is the amplitude of the exciton-phonon interaction. In the adiabatic approximation, the intensity of the one-phonon photoluminescence band is determined by the expression with only diagonal ($n_2 = n_1$) summands in (47):

$$F^{(\lambda)}_{n_0,1} = \sum_{n_1} \frac{\hbar^{-2}|f_{n_1}<n_1|\gamma_\lambda|n_0>|^2}{(\Omega_{n_0} - \Omega_{n_1} - \omega_\lambda)^2}. \tag{48}$$

We consider the Hamiltonian of the electron(hole)-phonon interaction in the form [42]

$$\hat{H}_{e(h)-ph} = \sum_\lambda \Gamma^{e(h)}_\lambda(\vec{r}_{e(h)})(a_\lambda + a^\dagger_{-\lambda}), \tag{49}$$

so that the amplitude of the exciton-phonon interaction $\gamma_\lambda$ in (47) and (48) is

$$\gamma_\lambda = \Gamma^e_\lambda(\vec{r}_e) - \Gamma^h_\lambda(\vec{r}_h). \tag{50}$$

The oscillator strength (41) determines not only the photoluminescence spectrum, but also the radiative decay time $\tau_n$ for the exciton state with quantum number $n$ [31]:

$$\tau_n = \frac{2\pi\varepsilon_0 m_0 c^3 \hbar^2}{\kappa e^2 \left[E^n_{exc}(d_1)\right]^2 f_n(d_1)}, \tag{51}$$

where $\kappa$ is the refractive index.

**VI. Results and comparison of the theory with experiment**

The developed method for the calculation of the exciton states in $Al_xGa_{1-x}N/GaN$ QW heterostructures is applicable for all values of $x$.

Photoluminescence spectra in $Al_xGa_{1-x}N/GaN$ heterostructures have been observed in Refs. [7,9,11,12,14,16,21]. In MQW heterostructures with $x=0.17$ [14], distinct zero-phonon and one-phonon peaks have been detected. The oscillator strength of the one-phonon peak is ~10% of that of the zero-phonon one. An heterostructure with $x=0.24$ [10] is characterized by a strong built-in electrostatic field ($F$~1.5MV/cm). The numerical calculations are performed with the geometric and material parameters of the heterostructures, given in Refs. [10,14,35]. In both cases, $x=0.17$ and $x=0.24$, the barriers are sufficiently high and wide, so that one can neglect the overlap of the electron (or hole) wave functions from different QWs in a MQW heterostructure. An indirect link between the states of the charge carriers in different QWs exists. Indeed, the built-in field in each well and in each barrier is calculated using the parameters of the whole structure, see (6) and (7). The values of the parameter $F_0$, introduced in Sec. II, are 2.05 MV/cm for an heterostructure with $x=0.24$ [10] and 816.5 kV/cm for an heterostructure with $x=0.17$ [14].

In Fig. 1, graphs of the electron wave functions in a QW with width $d_1=12$ ML=3.108 nm are shown for a MQW heterostructure with four QWs [14] and barriers with thickness $d_2=30$ nm. In contrast to the symmetrical electron wave function in the rectangular QW, the wave function $\phi_e(x,y,z)$ in the triangular QW is asymmetrical: it possesses a maximum at $z\approx 3$ ML and smoothly decreases with increasing $z$ at $z>3$ ML.

The five lowest exciton energy levels in a QW with width $d_1=16$ ML=4.144 nm for a MQW heterostructure $Al_{0.17}Ga_{0.83}N/GaN$, which contains four wells and barriers with thickness $d_2=30$ nm [14], are shown in Fig. 2. In our calculations of the exciton states we used 80 ($I=1$, $J=10$, $K=8$) basis functions. In order to estimate the precision of the calculation, we computed exciton energy spectra also with 128 ($I=1$, $J=16$, $K=8$) and 256 ($I=2$, $J=16$, $K=8$) basis functions. It is found that a basis of 80 functions provides a precision better than 0.5 meV for the calculation of the 40 lowest exciton energy levels.

Comparison of the calculated- and the experimentally obtained dependencies of the exciton transition energies on the width of the QW in the MQW heterostructures $Al_{0.17}Ga_{0.83}N/GaN$, with different barrier-widths, is presented in Fig. 3. In the insert, we show the deviations of the transition energies, calculated in the present work, from the transition energy obtained by means of a variational approach in Ref. [14]. The same parameter $F_0=816.5$ kV/cm is used for all curves. Good agreement is found between the calculated exciton transition energies and experiment. Another comparison of theoretical and experimental [10] exciton transition energies as a function of the QW width for heterostructures with $x=0.24$ is shown in Fig. 4. In this case, there is agreement with experiment for all investigated widths of the QWs. In the insert, we show the deviations $E_{exc}^n - E_{variational}^n$ of the transition energies $E_{exc}^n$, calculated in the present work, from the transition energy $E_{variational}^n$ found using a variational approach in Ref. [10].

Results of the calculations of the oscillator strengths, using (41) to (43), for an $Al_{0.24}Ga_{0.76}N$/GaN MQW heterostructure with $d_1$=3 nm and $d_2$=5 nm for the cases $(\vec{e} \perp c)$ and $(\vec{e} \| c)$ are shown, respectively, in Fig. 5(a) and Fig. 5(b). In the inserts, the oscillator strengths are shown in the absence of the built-in field. In the case of the in-plane $(\vec{e} \perp c)$ electric field, apart from the peaks at low exciton energies, pronounced peaks of the oscillator strength occur at higher energies. The appearance of those peaks is explained by the increase of the overlap of the hole size-quantized wave functions with the electron ground-state wave function for higher levels of the size-quantized holes. This situation is illustrated in Fig. 6. As seen from this figure, the overlap between the hole exited-state wave function and the electron ground-state wave function is stronger than the overlap between the hole ground-state wave function and the electron ground-state wave function.

The radiative decay time $\tau_n(d_1)$, calculated using (51), is presented in Fig. 7. The experimentally obtained values of $\tau_n$ are shown by squares. Our theoretical curve is in closer agreement with the experimental points than the theoretical curve presented in Ref. [10]. Our estimated built-in field is close to the fitting field from Ref. [10]. From Fig. 7 it follows that for small QW widths ($d$ < 1.5 nm) the effect of the built-in field on the radiative decay time is significant only at sufficiently large values $F$ > 1.2 MV/cm.

Our calculation shows that for the heterostructure under consideration, only the interface and bulk-like optical phonons determine the oscillator strength of the phonon sideband of the photoluminescence. The phonon energies $\hbar\omega(q)$ of the interface and bulk-like modes in the $Al_xGa_{1-x}N$/GaN heterostructure calculated according to Ref. [42] are presented in Fig. 8. Using Eqs. (45) to (47), the photoluminescence band is calculated including the zero-phonon and one-phonon peaks for the $Al_{0.17}Ga_{0.83}N/GaN$ MWQ heterostructure containing QWs with width $d_1$=16 ML (see Fig. 9). The one-phonon peak is shown in the insert. Its oscillator strength, obtained with the non-adiabatic theory, is one order of magnitude larger than that calculated in the adiabatic approximation [cf. Eq. (48)]. The position of the zero-phonon peak and the ratio of the oscillator strengths of the one-phonon and zero-phonon peaks derived with our theory are in a fair agreement with experiment.

**VII. Conclusions**

A theory of the exciton states in planar $Al_xGa_{1-x}N/GaN$ heterostructures is developed, using the 6-band hole model with an accurate expression for the electron-hole interaction. For the first time, the exciton energy spectrum in such heterostructures is obtained, including as many as 40 excited states. When calculating the photoluminescence spectra, optical phonons specific for wurtzite crystals are taken into consideration. Comparison of our results with those obtained using a variational single-band approximation indicates that the applicability of the latter is limited to the calculation of the exciton ground state in relatively thick layers.

The observed optical properties of the heterostructure under analysis are sensitive to the built-in electrostatic fields, induced by the piezoelectric and spontaneous polarizations. Those electrostatic fields turn the blue shift of the photoluminescence band, due to size quantization with increasing width of the QWs, into a red shift. This effect is enhanced by the increase of the built-in field. The frequency of the photoluminescence band as a function of the QW width is well described by the present theory. The built-in field leads to a decreasing overlap of the electron and hole wave functions and, consequently, to an increase of the radiative decay time $\tau_n$. The increase of $\tau_n(d_1)$ with increasing $d_1$ is well described by the present theory for all experimentally available values of $d_1$.

Finally, we have demonstrated that a non-adiabatic approach is needed in order to quantitatively interpret the observed positions and the ratios of the intensities of the one-phonon and zero-phonon photoluminescence peaks in the wurtzite $Al_xGa_{1-x}N$/GaN QW heterostructures.

**Acknowledgements**

This work was supported by FWO-V project G.0435.03, the WOG WO.035.04N, GOA BOF UA 2000 and IUAP (Belgium), the European Commission SANDiE Network of Excellence, contract no. NMP4-CT-2004-500101 and INTAS project no. 05-104-7656. The work in the State University of Moldova was supported in part by the U.S. Civil Research and Development Foundation, project MOE2-3057-CS-03.

**Figure Captions**

Fig. 1. Ground state electron wave function in rectangular ($F = 0$) and triangular ($F = 780$ kV/cm) quantum wells. The built-in electric field $F = 780$ kV/cm corresponds to the value $F_0$=816.5 kV/cm, used in our calculations.

Fig. 2. The five lowest exciton energy levels in the Al$_{0.17}$Ga$_{0.83}$N/GaN MQW heterostructure of Ref. [14] containing four quantum wells with $d_1$ =16 ML (1 ML=0.259 nm) and barriers with $d_2$ = 30 nm.

Fig. 3. Well-width dependence of the calculated exciton transition energies in an Al$_{0.17}$Ga$_{0.83}$N/GaN MQW heterostructure and the experimental peak positions from Ref. [14]. Insert: deviations $\Delta$ of the transition energies, calculated in the present work, from the transition energy found by means of a variational approach in Ref. [14].

Fig. 4. Well-width dependence of the calculated exciton transition energies in an Al$_{0.24}$Ga$_{0.76}$N/GaN heterostructure and the experimental peak positions from Ref. [10]. Insert: deviations $\Delta$ of the present results from those calculated in Ref. [10].

Fig. 5. Exciton oscillator strengths as a function of the exciton energy in an Al$_{0.24}$Ga$_{0.76}$N/GaN heterostructure containing a quantum well with $d_1 = 3$ nm and barriers with $d_2 = 5$ nm for two polarizations of light: $\vec{e} \perp c$ (panel a) and $\vec{e} \parallel c$ (panel b). The built-in electrostatic field $F$ is 1.48 MV/cm for both cases (a) and (b). Inserts: the oscillator strengths for the case $F = 0$.

Fig. 6. The electron wave function $\phi_e^{i=1}(z)$ and the components $\Upsilon_2^{j=1}(z)$ and $\Upsilon_2^{j=7}(z)$ of the hole wave functions.

Fig. 7. Photoluminescence decay time as a function of the QW thickness for the Al$_{0.24}$Ga$_{0.76}$N/GaN heterostructure with 5-nm barriers for different values of the built-in electrostatic field. The experimental points of Ref. [10] are shown as filled squares.

Fig. 8. Interface and confined optical phonon modes in an Al$_{0.17}$Ga$_{0.83}$N/GaN heterostructure containing a QW with $d_1$ =16 ML and barriers of infinite thickness.

Fig. 9. Photoluminescence spectra of an Al$_{0.17}$Ga$_{0.83}$N/GaN MQW heterostructure with four 16-ML QWs and 30-nm barriers. The positions of the photoluminescence peaks and their relative intensities are in a fair agreement with experiment of Ref. [14].

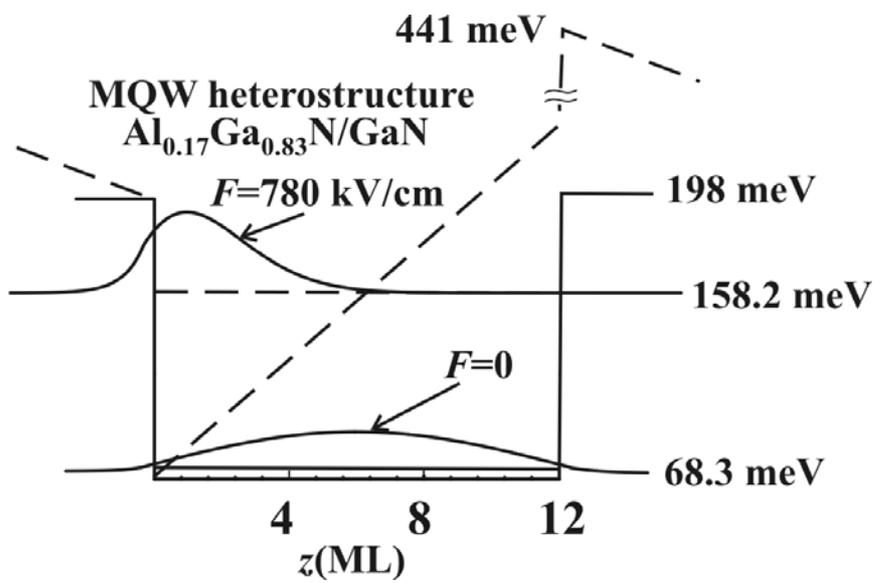

Fig. 1. E. P. Pokatilov, D. L. Nika, V. M. Fomin and J. T. Devreese.

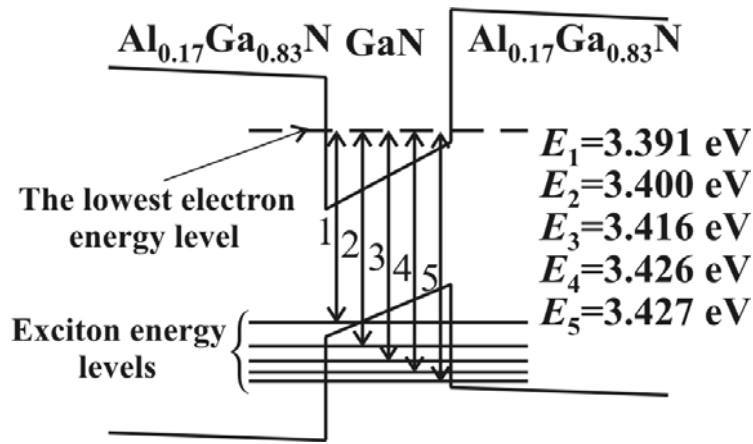

Fig. 2. E. P. Pokatilov, D. L. Nika, V. M.Fomin and J. T. Devreese.

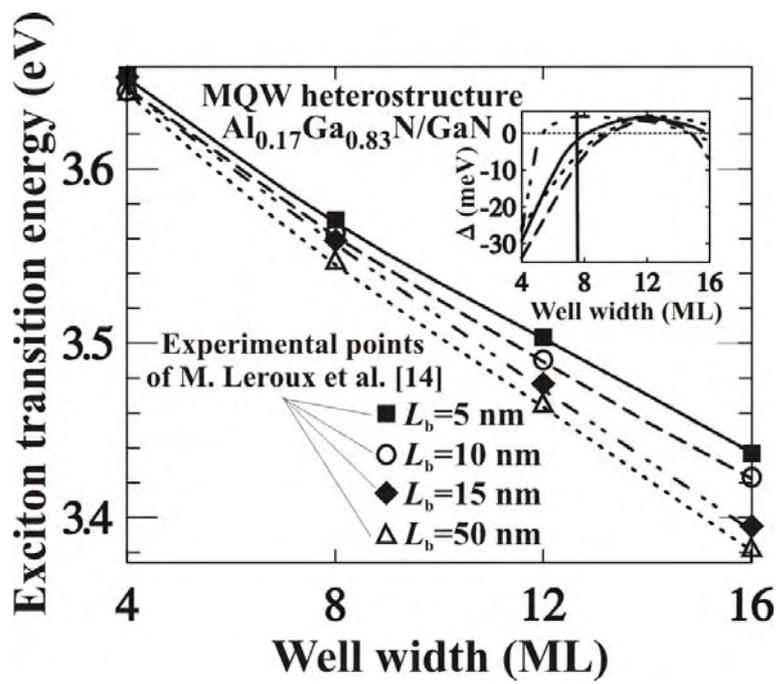

Fig. 3. E. P. Pokatilov, D. L. Nika, V. M. Fomin and J. T. Devreese.

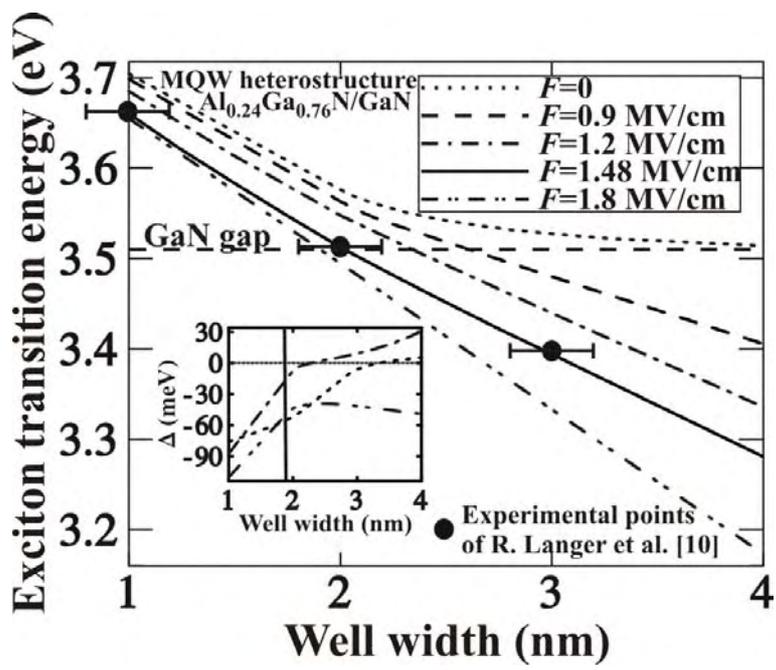

Fig. 4. E. P. Pokatilov, D. L. Nika, V. M.Fomin and J. T. Devreese.

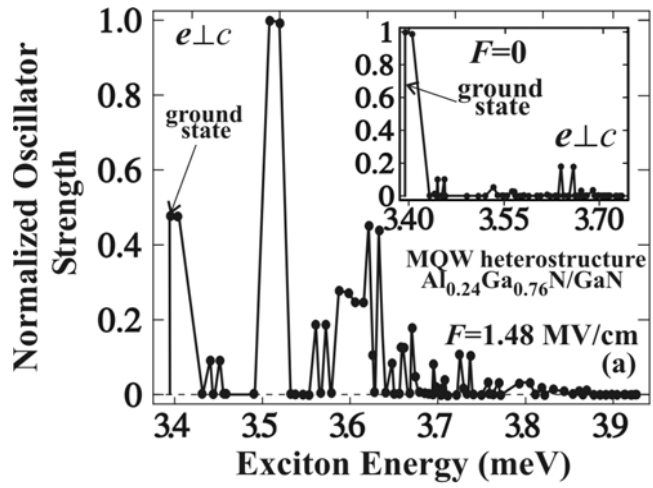

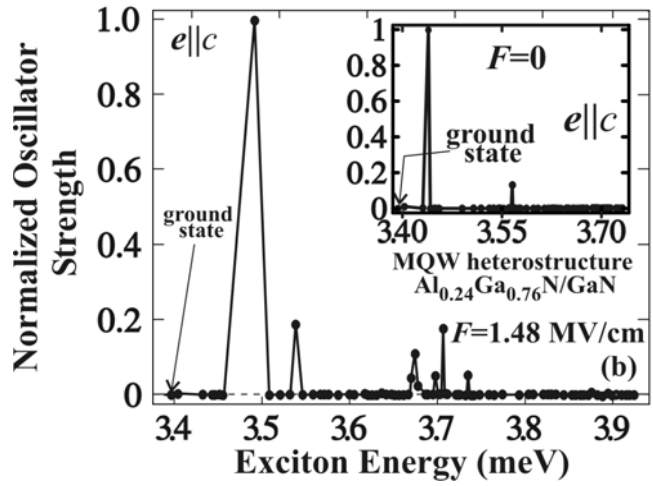

Fig. 5. E. P. Pokatilov, D. L. Nika, V. M.Fomin and J. T. Devreese.

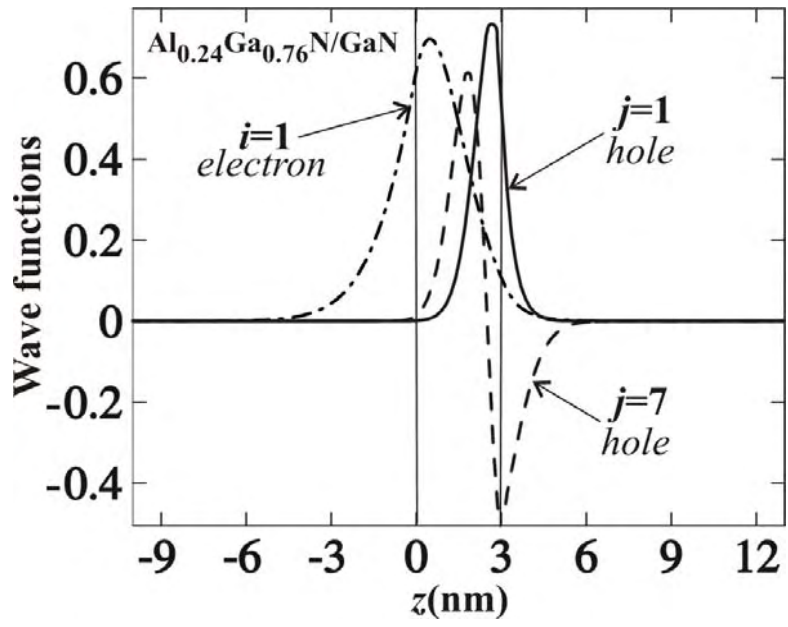

Fig. 6. E. P. Pokatilov, D. L. Nika, V. M.Fomin and J. T. Devreese.

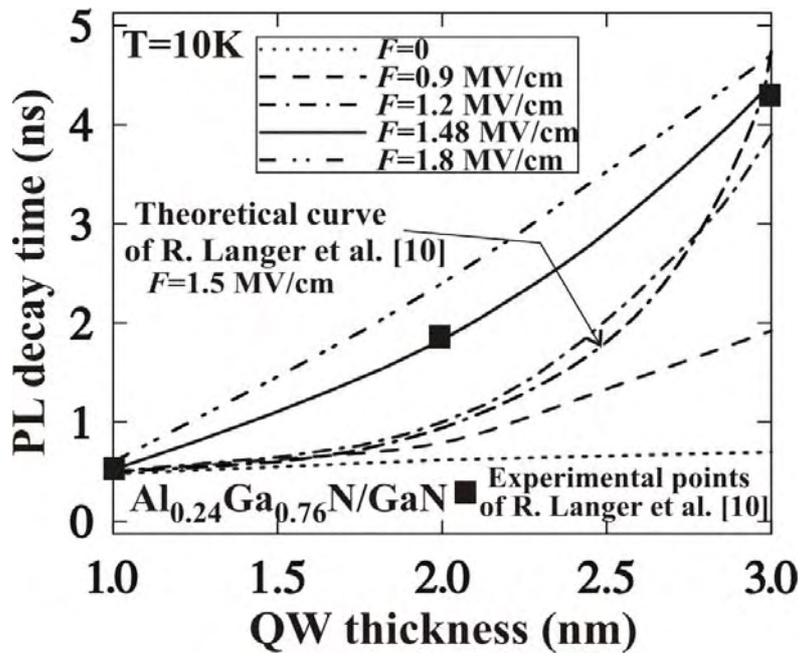

Fig. 7. E. P. Pokatilov, D. L. Nika, V. M.Fomin and J. T. Devreese.

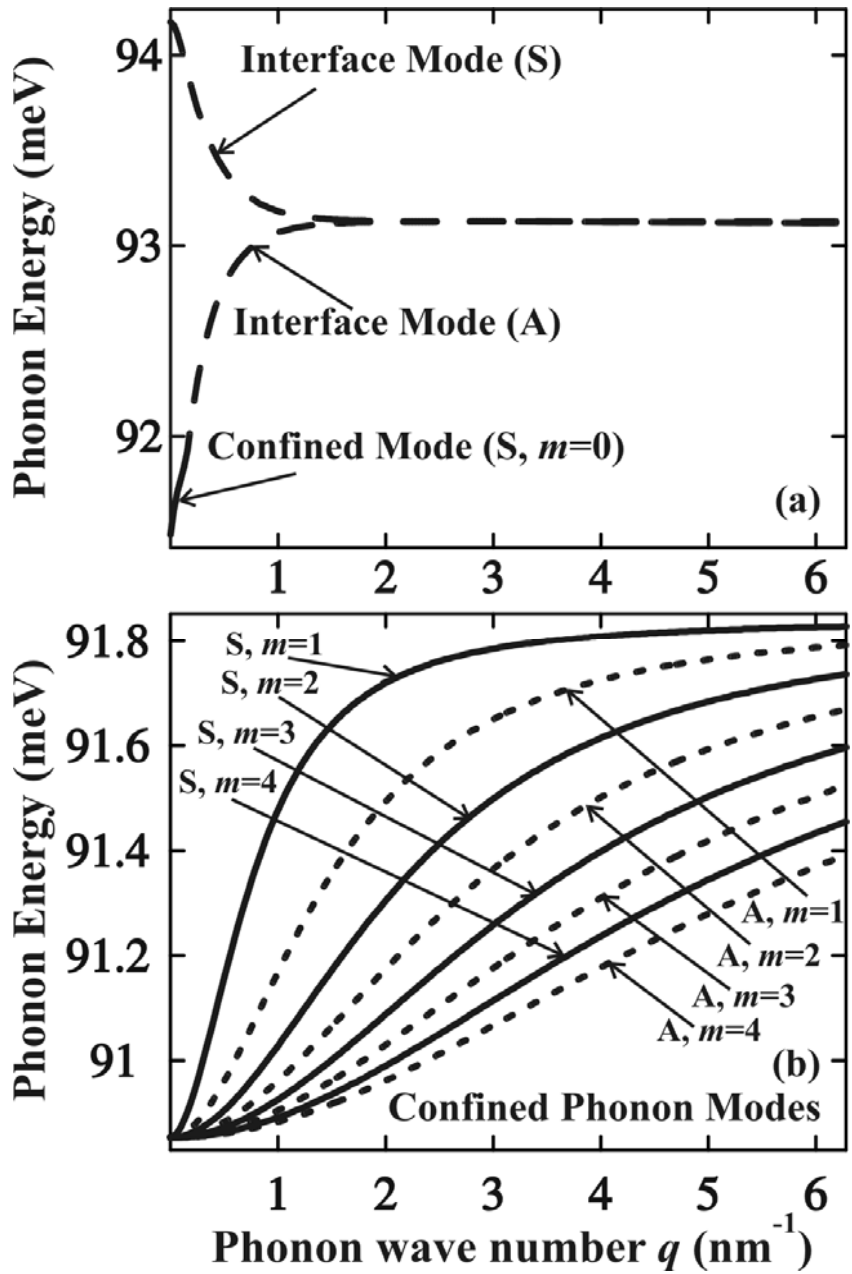

Fig. 8. E. P. Pokatilov, D. L. Nika, V. M.Fomin and J. T. Devreese.

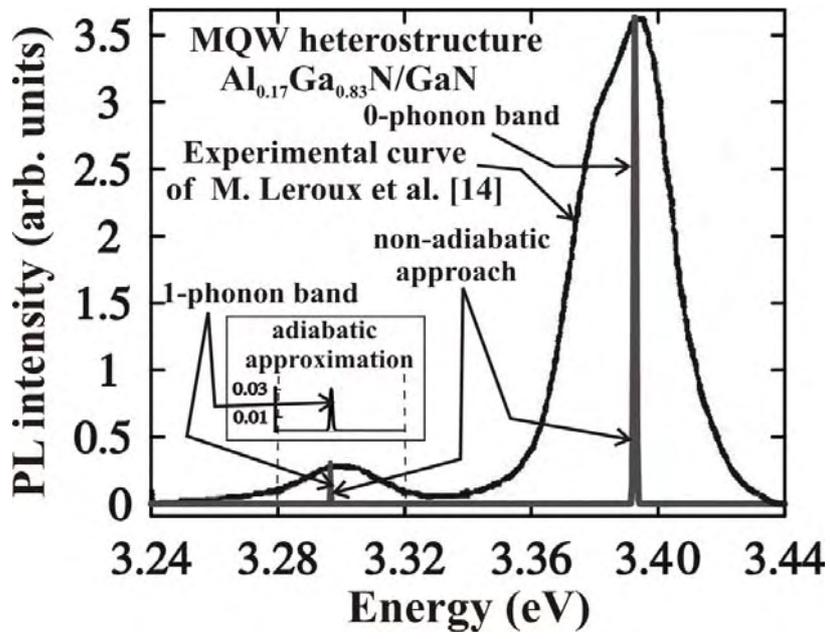

Fig. 9. E. P. Pokatilov, D. L. Nika, V. M.Fomin and J. T. Devreese.